\documentstyle[epsfig,epsf]{aipproc}
\newcommand{\DO}{D\raise1pt\hbox{$\not$}O}
\begin{document}
\title{
Photon and Pion Production at High $p_T$}
\author{
Marek~Zieli\'{n}ski
}
\address{
\centerline{University of Rochester, Rochester, New York 14627}          
}
\maketitle
\begin{abstract}
We present a study of high-$p_T$ photon  and pion production in
hadronic interactions, focusing on a comparison of the 
yields with expectations from next-to-leading order
perturbative QCD (NLO pQCD). We examine the impact of 
phenomenological models of $k_T$ smearing (which approximate
effects of additional soft-gluon emission) on
absolute predictions for photon and pion production
and their ratio.

\end{abstract}

Single and double 
direct-photon production in hadronic collisions at high transverse
momenta ($p_T$) have long been viewed as an ideal testing ground for
the formalism of pQCD.  A reliable theoretical description of the
direct-photon process is of special importance because of its
sensitivity to the gluon distribution in a proton through the
quark--gluon scattering subprocess ($gq\rightarrow\gamma q$).  The
gluon distribution, $G(x)$, is relatively well constrained for $x<0.1$,
but much less so at larger~$x$~\cite{huston-uncertainty}.  In
principle, fixed-target direct-photon production can constrain $G(x)$
at large~$x$, and such data have therefore been incorporated in
several modern global parton distribution function (PDF)
analyses~\cite{cteq4,grv92,mrst}.

\begin{figure}
\vspace{-.6in}
\centerline{
\hspace{0.3in}
\epsfxsize=2.5truein
\epsffile{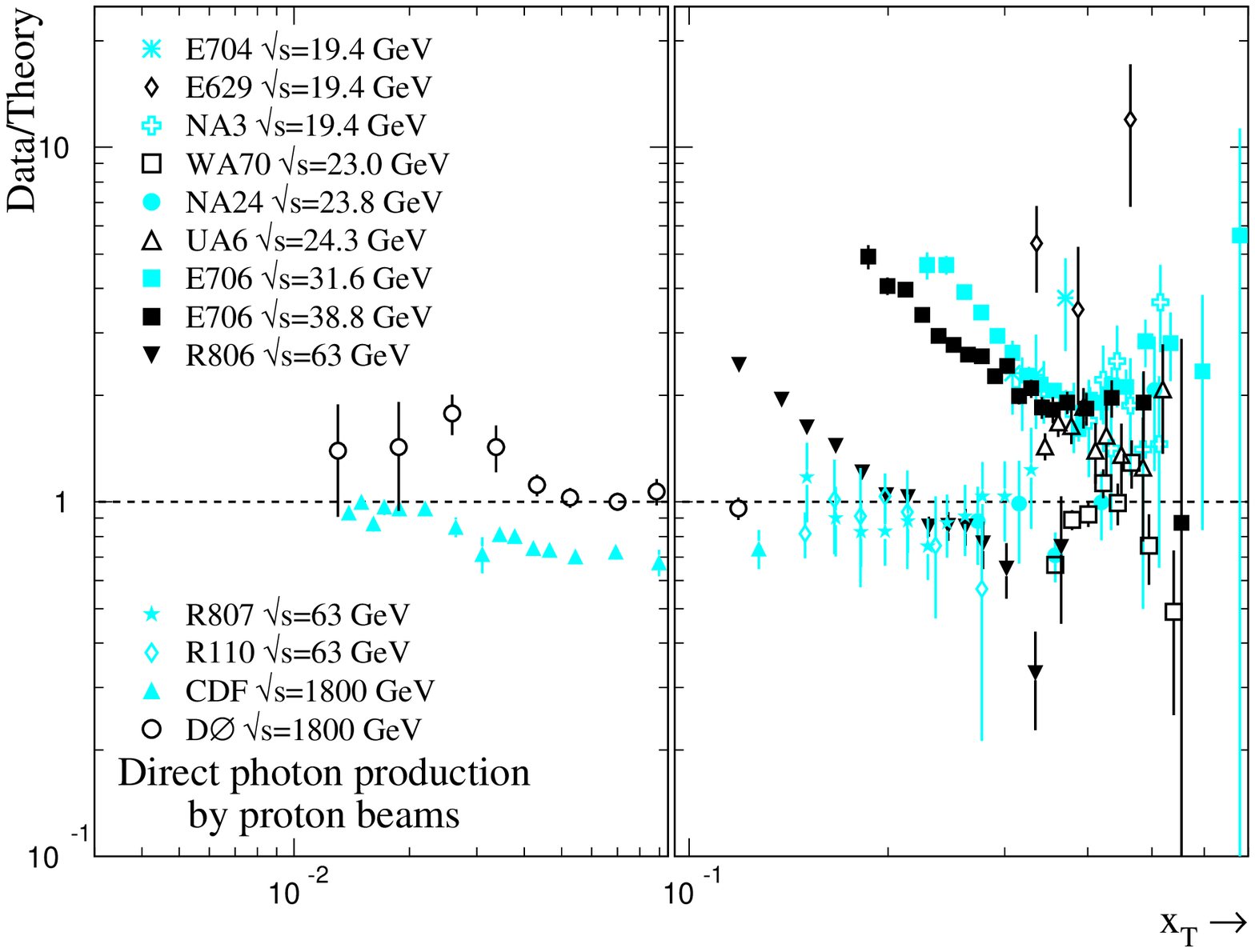} 
\hspace{0.3in}
\epsfxsize=2.5truein
\epsffile{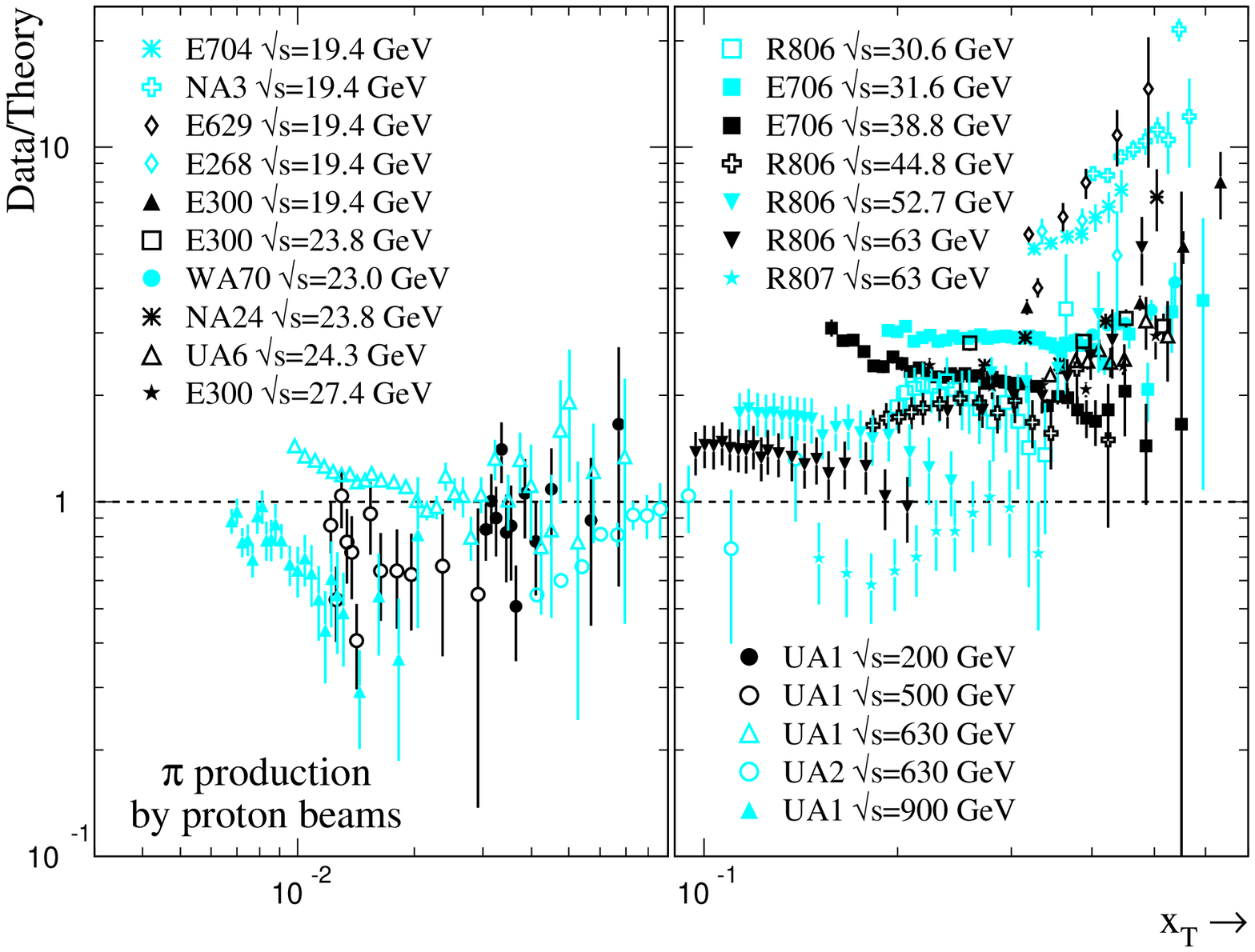}}
\vspace{-.7in}
\caption{Comparison of direct-photon (left)
and pion (right) data to NLO pQCD vs  $x_T$.
\label{fig:discrepancy}}
\end{figure}

However, both the completeness of the NLO description of
the direct-photon process, as well as the consistency of results from
different experiments, have been questioned~\cite{mrst,baerreno,%
huston-discrepancy,E706-kt,ktprd,aurenche-dp,aurenche-pi0,kimber}.
The inclusive production of hadrons provides a further means of
testing the predictions of the NLO pQCD formalism.  Deviations have
been observed between measured inclusive direct-photon and pion cross
sections and NLO pQCD calculations.  Examples of such discrepancies
are shown in Fig.~\ref{fig:discrepancy}
where ratios of data to theory are displayed as a function of
$x_T=2p_T/\sqrt{s}$ 
for photon and pion data.
(Unless otherwise indicated, all NLO
calculations~\cite{aurenche-nlo,qiu,gordon,aversa,bailey}
in this paper use a single scale of $\mu=p_T/2$, CTEQ4M
PDFs~\cite{cteq4}, and BKK fragmentation functions for
pions~\cite{BKK}.)  
It has been suggested that part of the deviations from theory for
both photons and pions can be ascribed to higher-order effects of
initial-state soft-gluon radiation~\cite{huston-discrepancy,E706-kt,ktprd}. 

\begin{figure}[t]
\vskip-.3in
\centerline{
\hspace{0.2in}
\epsfxsize=2.3truein
\epsffile{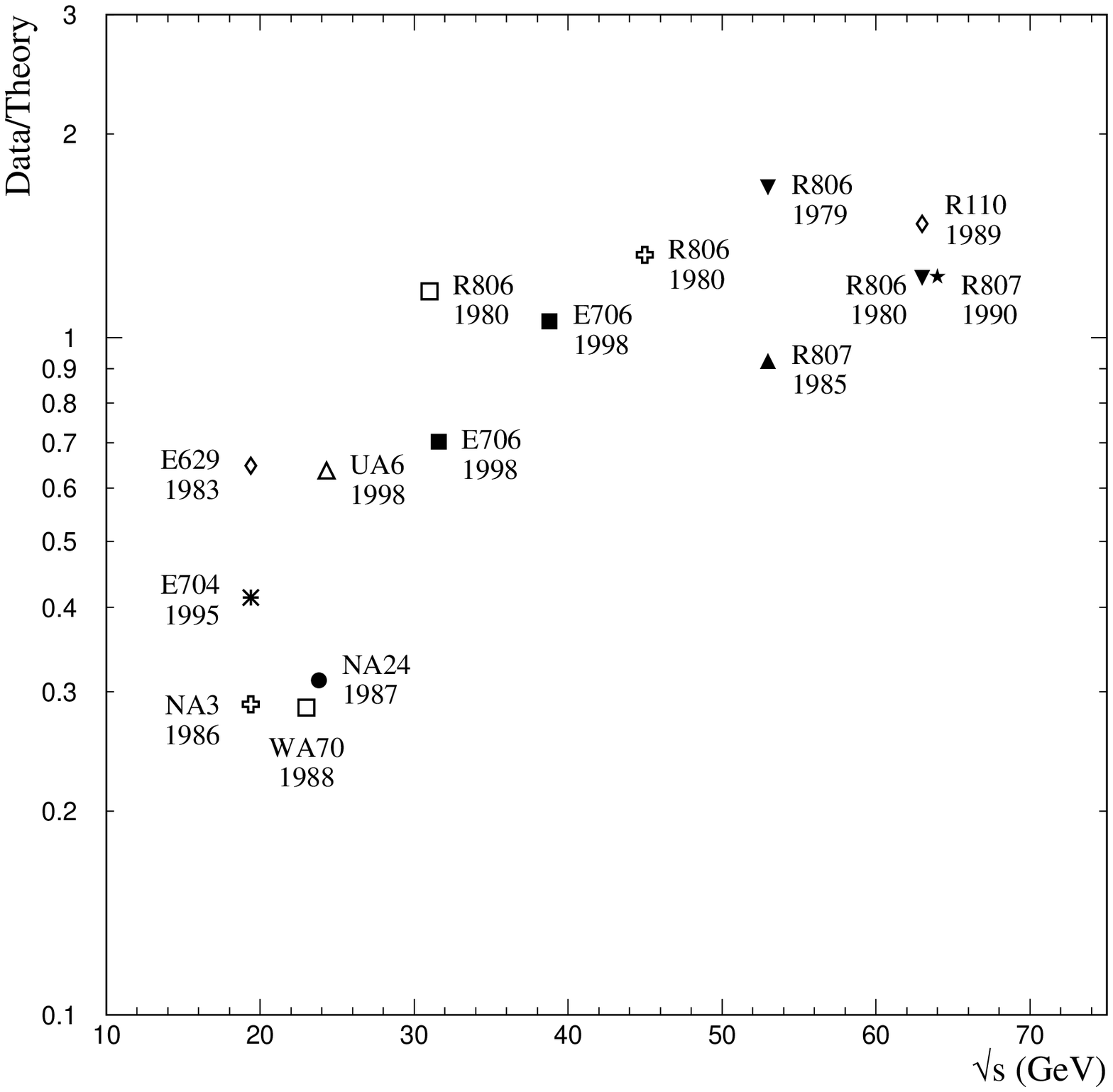}
\hglue1truept\hspace{0.7in}
\epsfxsize=2.3truein
\epsffile{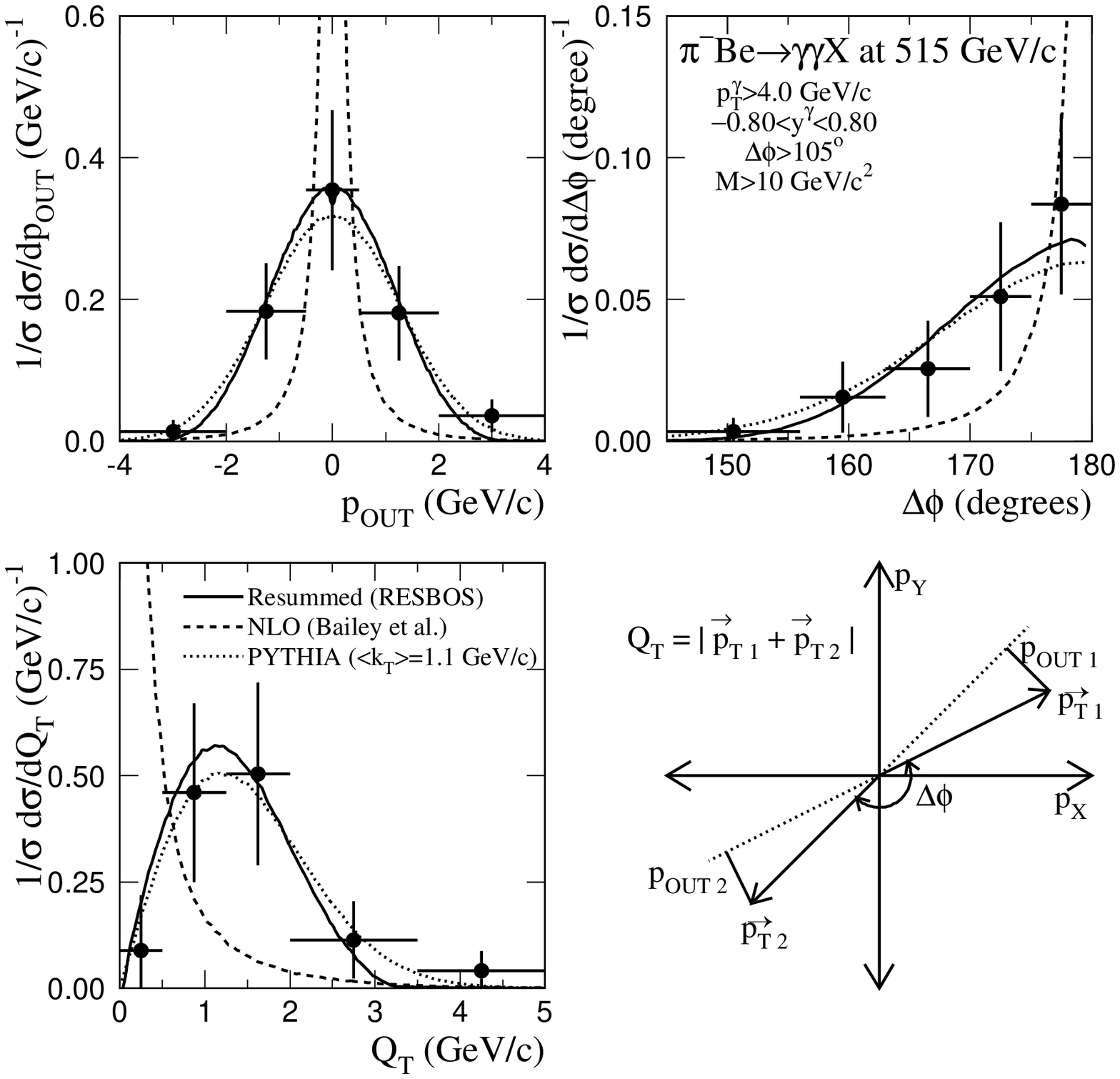}
}
\vspace{-.3in}
\caption{Left:
The ratio of data for the $\gamma/\pi^0$ measurements to NLO theory,  
as a function of $\sqrt{s}$.
Right: Diphoton data from E706~\protect\cite{begel} compared to 
NLO~\protect\cite{bailey} (dashed),
resummed~\protect\cite{RESBOS} (solid), and Pythia~\protect\cite{pythia61} 
(dotted) calculations, using GRV PDF~\protect\cite{grv92}.
\label{fig:photonTOpi0_gg}}
\end{figure}

Given the scatter of the data 
shown in Figs.~\ref{fig:discrepancy}, 
it may be instructive to consider measurements
of the $\gamma/\pi^0$ ratio 
over a wide range of $\sqrt{s}$~\cite{osaka}.  
Both experimental and theoretical uncertainties tend
to cancel in such a ratio, and the ratio should also be less
sensitive to incomplete treatment of gluon radiation. 
A compilation of comparisons between data and theory, shown for 
simplicity without their
uncertainties, is presented in Fig.~\ref{fig:photonTOpi0_gg}.  
In
this figure, the ratios of data to theory for the $\gamma$ to $\pi^0$
measurements have been approximated as a constant value at high-$p_T$, 
and the results plotted as a function of $\sqrt{s}$ 
(see \cite{osaka} for details).  The figure suggests an
energy dependence in the ratio of data to theory for $\gamma/\pi^0$
production.  There is, however, an indication of substantial
differences between the experiments at low $\sqrt{s}$ (where the
observed $\gamma/\pi^0$ is smallest), which makes it difficult to
quantify this trend.  Recognizing the presence of these differences is
especially important because thus far only the low energy photon experiments
have been used in PDF fits to extract the gluon distribution.

The differences between many of the data sets and pQCD, seen in 
Fig.~\ref{fig:discrepancy}, 
may be due to the impact of the effective parton transverse momentum, $k_T$.
In hadronic hard-scattering processes, there is generally a
substantial amount of $k_T$ in
the initial state resulting from gluon emission~\cite{ktprd}.
The presence of $k_T$ impacts the final state and has been
observed in measurements of Drell-Yan, diphoton, and heavy quark
production;  the 
amount of $k_T$ expected from NLO calculations is not sufficient 
to describe the data. The effective
values of $\langle k_T\rangle$/parton for these
processes vary from $\approx 1$ GeV/$c$ at fixed target energies,
as illustrated in Fig.~\ref{fig:photonTOpi0_gg} 
for diphoton distributions from E706~\cite{begel}, to 3--4 GeV/$c$
at the Tevatron Collider --- the growth is approximately logarithmic with
center-of-mass energy~\cite{ktprd}.  
The size of the $\langle
k_T\rangle$ values, and their dependence on energy, argue against a
purely ``intrinsic'' non-perturbative origin. Rather, the major part
of this effect is generally attributed to soft-gluon emission.  While
the importance of including gluon emission through the resummation
formalism has long been recognized and calculations have been
available for some time for Drell-Yan~\cite{altarelli},
diphoton~\cite{RESBOS,fergani}, and W/Z production~\cite{RESBOS}, they
have only recently been developed for inclusive direct-photon
production~\cite{nason,kidonakisowens,laenen,lilai,li,sterman}.

In the absence of a rigorous theoretical treatment of the impact
of gluon emission on high-$p_T$ inclusive production, a more
intuitive phenomenological approach has proved successful~\cite{ktprd}.  
The soft gluon radiation was parametrized in terms of an effective
$\langle k_T\rangle$ that provided an additional transverse impulse to
the outgoing partons. Because of the steeply falling cross section in
$p_T$, such a $\langle k_T\rangle$ can shift the production of
final-state particles from lower to higher values of $p_T$,
effectively enhancing the cross section.

\begin{figure}
\centering\leavevmode
\vglue1truept\vspace{-0.25in}
\hglue1truept\hspace{-3in}
\epsfxsize=2.3truein
\epsffile{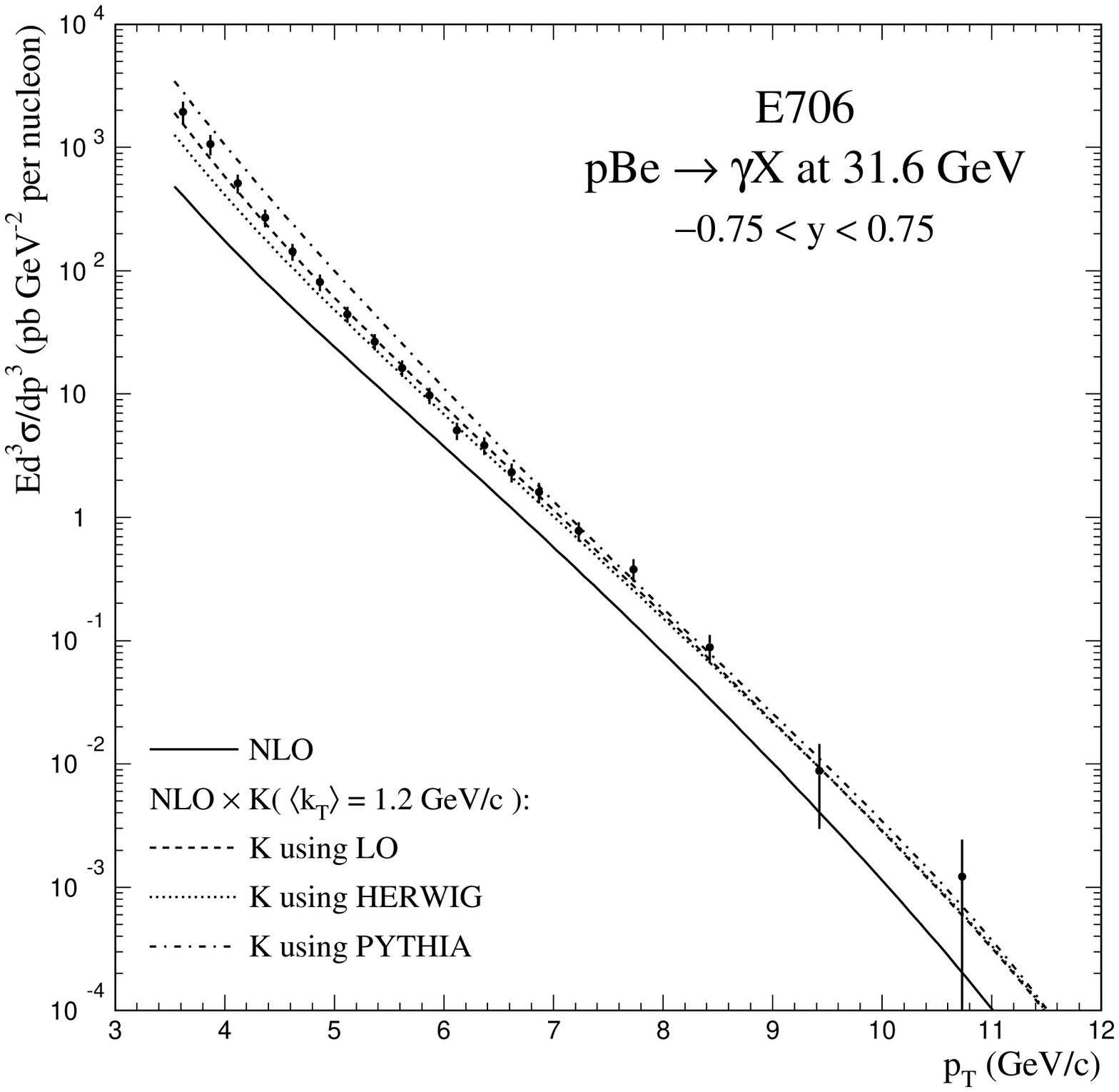}
\vskip -7.5cm
\hglue1truept\hskip 2.5in
\epsfig{figure=Zielinski-Parallel4.w1.2-e706crs.eps,height=2.3truein,angle=90}
\vskip 1cm
\vspace{-.3in}
\caption{Left:
Comparison between the E706 direct-photon data at 
$\sqrt{s}=31.6$~GeV~\protect\cite{E706-kt} 
and the NLO pQCD calculation (solid), and the NLO
theory enhanced by K-factors obtained using the LO
calculation~\protect\cite{owens}
(dashed), {\sc herwig}~\protect\cite{herwig61} (dotted), and 
{\sc pythia}~\protect\cite{pythia61} (dash-dotted).
Right: 
Same data compared to recent QCD calculations.  The dotted
line represents the full NLO calculation~\protect\cite{gordon}, while
the dashed and solid lines, respectively, incorporate purely threshold
resummation~\protect\cite{nason} and joint threshold and recoil
resummation~\protect\cite{sterman}.
\label{fig:sterman}}
\end{figure}

As described in~\cite{ktprd}, a leading-order (LO) pQCD
calculation~\cite{owens} has been used to generate K-factors
(ratios of calculations for any given $\langle k_T\rangle$ to the
result for $\langle k_T\rangle=0$) for inclusive cross sections.
These $p_T$-dependent factors have been then
applied to the NLO pQCD calculations.  
The enhancements that would be expected for
direct-photon production from parton-showering
models~\cite{pythia61,herwig61} have also been investigated~\cite{osaka}. 
These programs do not provide
sufficient smearing at fixed-target energies because shower
development is constrained by cut-off parameters that ensure the
perturbative nature of the process.  Consequently, these calculations
allow additional input $k_T$ for Gaussian smearing, and are often used
that way in comparisons to data.
The respective corrections have been obtained 
using default settings for other program
parameters and an input $\langle k_T\rangle$ of 1.2 GeV/$c$ 
for the smearing,
relative to these same settings with $\langle k_T\rangle=0$, 
and then applied
to NLO pQCD calculations. The resulting comparisons to
data from E706~\cite{E706-kt} are displayed in Fig.~\ref{fig:sterman}
(similar results hold for pion production, not shown).
The observed differences should be kept in mind
when comparing these models to data for $k_T$-sensitive 
quantities.  

Recently, there has been significant progress in more rigorous
resummed pQCD calculations for single direct-photon production 
~\cite{nason,kidonakisowens,laenen,lilai,li,sterman}.
Substantial corrections to fixed-order QCD calculations are expected
from soft-gluon emission, especially in regions of phase space where
gluon emission is restricted kinematically. 
At large~$x$, there is a
suppression of gluon radiation due to the rapidly falling
parton distributions and a complete description of the cross section in
this region requires the resummation of ``threshold'' terms.  
Two recent threshold-resummed pQCD calculations for direct
photons~\cite{nason,kidonakisowens} exhibit far less dependence on QCD
scales than found in NLO theory.  These calculations agree with the
NLO prediction for the scale $\mu\approx p_T/2$ at low $p_T$ (without
inclusion of explicit $k_T$ or recoil effects), and show an
enhancement in cross section at high~$p_T$.

A method for simultaneous treatment of recoil and threshold
corrections in inclusive single-photon cross sections is being
developed~\cite{sterman}
within the formalism of collinear factorization.  
This approach accounts explicitly for the recoil from
soft radiation in the hard-scattering subprocess, and conserves both
energy and transverse momentum for the resummed radiation.  The
possibility of substantial enhancements from higher-order perturbative
and power-law nonperturbative corrections relative to NLO are
indicated at both moderate and high $p_T$ for fixed-target energies,
similar to the enhancements obtained with the simple $k_T$-smearing
model discussed above.
Figure~\ref{fig:sterman} (right) displays the results of an example
calculation~\cite{sterman} based on this approach 
compared with direct-photon measurements from E706.  

While there is still no resummation calculation for inclusive pion
production, the trend of recent developments in direct-photon
processes has led to an increased appreciation of the importance of
the effects of multiple gluon emission, and to the emergence of tools
for incorporating these effects.  These latest theoretical
developments encourage optimism that the long-standing difficulties in
developing an adequate description of these processes can eventually
be resolved, making possible a global re-examination of parton
distributions with an emphasis on the determination of the gluon
distribution from the direct-photon data~\cite{sterman-pdf}.

\acknowledgments

This work has been done in collaboration with L. Apanasevich, 
M.~Begel, C.~Bromberg, T.~Ferbel,  G.~Ginther,
J.~Huston, S.~Kuhlmann, P.~Slattery, and V.~Zutshi.
We also wish to thank  
P.~Aurenche, C.~Bal\'{a}zs, S.~Catani, M.~Fontannaz,
N.~Kidonakis, J.~F.~Owens, E.~Pilon, G.~Sterman, 
W.~Vogelsang and J.~Womersley 
for helpful discussions.

\renewcommand{\baselinestretch}{1.}

\end{document}